# Thermodynamics of nuclei in thermal contact

Karl-Heinz Schmidt, Beatriz Jurado

CENBG, CNRS/IN2P3, Chemin du Solarium B.P. 120, 33175 Gradignan, France

**Abstract:** The behaviour of a di-nuclear system in the regime of strong pairing correlations is studied with the methods of statistical mechanics. It is shown that the thermal averaging is strong enough to assure the application of thermodynamical methods to the energy exchange between the two nuclei in contact. In particular, thermal averaging justifies the definition of a nuclear temperature.

**Introduction**

It has been discussed that two moderately excited nuclei, brought into contact, form a system with very peculiar thermodynamical properties [1]. Recent experiments revealed that the nuclear level density $\rho$ up to excitation energies $E$ of about 10 or even 20 MeV [2, 3] are well represented by a constant-temperature formula

$$\rho(E) \propto \exp\left(\frac{E}{T}\right) \qquad (1)$$

with a parameter $T$, which essentially does not depend on energy.

This behaviour has been explained by the gradual melting of Cooper pairs in the energy domain governed by strong pairing correlations [4].

According to the definition of temperature $\Theta$ in a microcanonical ensemble

$$\Theta = \left(\frac{\partial(\ln \rho)}{\partial E}\right)^{-1} \qquad (2)$$

the $T$ parameter of the nuclear level density can rather directly be interpreted as the thermodynamical temperature of the nucleus. Although in a strict sense the state density differs from the level density by the spin degeneracy of the nuclear levels, the slow increase of the spin cut-off parameter with energy has practically no influence on the energy dependence of the nuclear state density. Thus, in a restricted energy range, the spin degeneracy can well be considered by a constant factor [5].

When two nuclei are brought into thermal contact with low relative velocity, their evolution towards thermodynamical equilibrium may be considered. The relative velocity is assumed to be so low that heating of the two nuclei by one-body dissipation according to the window formula [6] is negligible. This is the case when the mean kinetic energy of a nucleon in one of the nuclei counted in the frame of the other nucleus is small compared to the temperatures involved. This scenario is assumed to be realized in nuclear fission before scission. However, it is usually not realized in heavy-ion collisions due to the higher relative velocity of projectile and target.

If the two nuclei, respectively nascent fission fragments, that are considered have different thermodynamical temperatures, it has been derived that all excitation energy is transferred to the nucleus with the lower temperature in a process of energy sorting [1]. This process is driven by entropy, which evolves towards its maximum possible value in an isolated system according to the Second Law of thermodynamics.

**Problems of size**

It is obvious that the system consisting of the two nuclei in contact forms a microcanonical ensemble, e.g. there is no heat bath. In addition, the total energy of the system is finite and a constant of time. The same is true for the total number of particles. Due to its small size, however, the temperature of an isolated nucleus is not well defined; it is subject to fluctuations. This becomes obvious by the fluctuations of the single-particle occupation distribution from one nuclear level to another. E.g., in the spirit of the exciton model [7], the excitation energy may be shared by a strongly different number of particles and holes. The analytical relations of the Fermi-Dirac statistics apply only to an average over the single-particle distributions of many neighbouring nuclear levels.

At low total excitation energy or close to the end of the energy-sorting process, there arise still some other difficulties: The nuclear levels are discrete, and they do not overlap any more. Thus, the nuclear level density cannot be expressed by a continuous function of energy. This puts the definition of the nuclear temperature, eq. (2), and its use for describing the thermodynamic behaviour of the system considered in doubt even more.

It is the aim of the present work to present a careful study of this problem, which was not explicitly mentioned in ref. [1], because the conclusions of the present work were already silently considered.

**Microscopic scenario**

In order to base our study on a solid and reliable foundation, we chose statistical mechanics on the microscopic level as the starting point of our study [8]. On this level, each one of the two nuclei in contact can be considered as a microcanonical object, which is not completely isolated but in contact with the other nucleus. In this sense, the other nucleus acts as a sort of heat bath. However, there are two important differences to a heat bath in canonical thermodynamics: First, the temperature of the other nucleus, acting as a kind of heat bath, fluctuates, as mentioned above, and, secondly, the energy reservoir of that nucleus is finite.

On the exact microscopic level, we may formulate the problem as follows: While the complete system, consisting of the two nuclei, is isolated, the two nuclei are allowed to exchange between themselves (i) energy e.g. by collision processes between nucleons in the contact region or coupling of the two nuclei to collective modes of the whole system, and, eventually, (ii) energy and nucleons, if transfer of protons and neutrons between the two nuclei is considered in addition.

If complete knowledge is assumed of all nuclear states and their single-particle configurations[a], it is possible to calculate the evolution of the system, allowing for energy transfer between the two nuclei, with the Monte-Carlo technique: Starting from an initial configuration, the system, consisting of the two nuclei in contact, may change to any other configuration, each one with the same probability, respecting the condition of fixed total energy. Most often, the changes of the configurations proceed in the two nuclei separately according to Bohr's compound-nucleus hypothesis [9]. However, the thermal contact allows also for a coupling. The amount of energy transfer in a collision between nucleons in the contact zone is limited by the condition that most single-particle occupation patterns are close to the Fermi distribution. There are only very few configurations with holes well below the Fermi level and particles well above, because this would imply that a large fraction of energy is stored in only few degrees of freedom. Again, the transition to every final state of the complete system which respects the conservation of total energy is equally probable. The average rate of such collisions is proportional to the thermal coupling between the two nuclei in contact. In nuclear fission, the thermal coupling would be realized by the neck region.

These considerations prove that the statistical evolution of the system is well defined, and it is even possible to predict it precisely with a realistic model. Even if the application of standard thermodynamical methods may be questionable, e.g. due to the fluctuation of the temperatures of the two nuclei involved as a function of energy partition, the application of statistical mechanics poses no problem at all. Thermal averaging or the thermodynamical limit are not required.

---
[a] In the following of this chapter, the explicit formulation is done in the independent-particle picture because it is more transparent. Inclusion of residual interactions, e.g. pairing correlations, would not change the final conclusions of the specific questions studied in this chapter.

Including the transfer of nucleons through the neck does not impose any additional difficulties: The configurations of the nuclei with one neutron or proton more, respectively less, should be included in the number of possible final configurations.

**Thermal averaging**

In the light of the scenario outlined in the previous chapter, the possible application of thermodynamics to the configuration of two nuclei in thermal contact may be reconsidered. The heat flow $dE/dt$ between macroscopic objects, which is a continuous quantity, is governed by the temperature difference $\Delta T = T_2 - T_1$ and the thermal resistance $R_T$ against energy exchange

$$\frac{dE}{dt} = \frac{\Delta T}{R_T} \quad . \tag{3}$$

In contrast, the energy exchange between the two nuclei in contact in the scenario considered above proceeds in steps of considerable magnitude. One nuclear collision in the contact zone or one nucleon transferred may increase or decrease the excitation energy of one and the other nucleus by an amount, which in most cases corresponds to the energy range where single-particle levels are partly filled. Due to pairing correlations, this is in the order of one MeV, even at low excitation energies. This energy is not much smaller than the total excitation energy of the system. Therefore, the process of energy exchange cannot be considered as a continuous process, but it proceeds in rather large steps.

At first sight, this seems to complicate the application of thermodynamics to the di-nuclear system even more, but the contrary is true: The transfer of energy between the two nuclei in steps causes an averaging of the thermodynamical properties of the two nuclei. After some steps of energy transfer, the thermal driving force corresponding to the entropy gradient averages over a considerable energy region. The width of this region is given by the magnitude of the energy steps. This thermal averaging smoothes out the fluctuations of the micro-canonical temperature defined by eq. (2) in the region where the nuclear level density is continuous but subject to fluctuations. Even more, in the region of discrete, not overlapping levels, this effect averages over the levels in a finite interval. Note that any of the discrete levels of the nucleus with the lower excitation energy is accessible if the partner nucleus with the higher excitation energy is in the regime of overlapping levels. Thus, the effective temperatures of the nuclei, which drive the process of energy exchange, can well be evaluated by eq. (2). However, instead of the true level density $\rho$, a smoothed level density $\bar{\rho}$ should be inserted into eq. (2), which results from averaging the measured nuclear level density $\rho$ over a finite energy interval. As a result of this averaging, the fluctuations and structures of the measured level densities are almost completely washed out, and the constant-temperature formula even better describes the effective averaged level density, which governs the energy exchange between nuclei in contact, in the excitation-energy domain of strong pairing correlations.

This thermal averaging is illustrated in Fig. 1, where the measured level density of $^{172}$Yb is compared with the result of a smoothing procedure. The smoothing procedure simulates the thermal averaging due to the finite magnitude of the energy transfer between two nuclei in thermal contact. Very clearly, the structures in the measured level density around 1.2 MeV and 2.4 MeV, corresponding to the first quasi-particle excitations, are completely washed out.

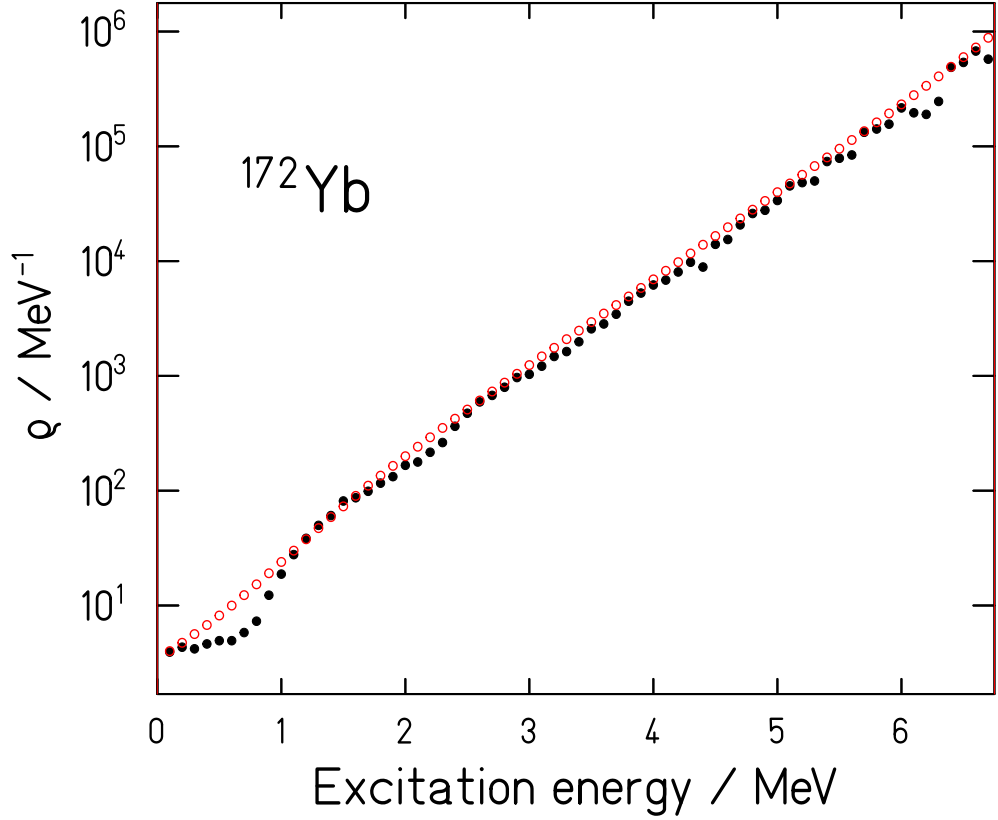

**Figure 1:** The measured level density [4] of $^{172}$Yb (full symbols) is compared with the result of a smoothing procedure (open symbols). The logarithm of the level density was convoluted with a Gaussian function with a standard deviation corresponding to the temperature $T$ of this nucleus. The value of $T = 0.57$ MeV was deduced from the level density around $E = 5$ MeV according to eq. (2). Border effects were avoided by extrapolating the logarithm of the level density to both sides, below 0.1 MeV and above 6.7 MeV, by a linear fit to the measured data before convolution.

Finally, we conclude that the process of energy transfer, respectively energy sorting, is governed by a differential equation similar to eq. (3), however, with an additional fluctuating term $F$.

$$\frac{dE}{dt} = \frac{\Delta T}{R_T} + F \qquad (4)$$

This fluctuating term induces a thermal averaging as described above and introduces a random behaviour of the energy transfer. It also prevents that the energy-sorting process finishes with a situation where the ground state of the nucleus with the higher temperature is populated with 100% probability. When transfer of nucleons is possible, also the even-odd mass differences have a sizeable influence on the last steps of the energy-sorting mechanism. The eventual exchange of one neutron and one proton may transform an odd-odd nucleus in its ground state to an even-even nucleus in its ground state, which is more strongly bound by about two times the pairing gap parameter $\Delta$. This leads to an increase in entropy due to the higher excitation energy available in the other nucleus. A schematic scenario, which explains the complex features of the even-odd effect in fission-fragment nuclear-charge distributions along these lines, has been described in ref. [10].

**Conclusions**

The recent discovery of the energy-sorting process in nuclear fission [1] was described using thermodynamical concepts and relations. In the present work, considerations based on statistical mechanics show that this description is justified in spite of the small size of nuclei. Indeed, the energy exchange between the two nuclei in contact proceeds in steps, e. g. by nuclear collisions in the contact zone or by nucleon transfer. The nucleons that take part in the collisions or in the transfer are those with energies close to the Fermi energy. Therefore, the energy transfer leads to a fluctuation of the excitation energy of one and the other nucleus by a considerable amount. After a number of steps, the system averages over a considerable energy region, and thus the entropy which determines the dynamical evolution of the system is an average value. Consequently, the irregularities of the level density are smoothed out by the very nature of the energy-exchange process, allowing for the definition of a nuclear temperature.

**Acknowledgment**

This work was supported by the EURATOM 6. Framework Program "European Facilities for Nuclear Data Measurements" (EFNUDAT), contract number FP6-036434.